\documentclass[12pt,preprint]{aastex}

\slugcomment{To appear in {\em The Astrophysical Journal}}

\shorttitle{Ray-tracing for complex astrophysical
high-opacity structures}
\shortauthors{Steinacker et al.}

\begin{document}


\title{Ray-tracing for complex astrophysical
high-opacity structures}


\author{J. Steinacker}
\affil{Max-Planck-Institut f\"ur Astronomie, K\"onigstuhl 17, D-69117 Heidelberg, Germany\\
Astronomisches Rechen-Institut am Zentrum f\"ur Astronomie Heidelberg,\\
M\"onchhofstr. 12-14, D-69120 Heidelberg, Germany}
\email{stein@mpia.de}

\author{A. Bacmann}
\affil{Observatoire de Bordeaux, 2 Rue de l'Observatoire, BP 89, 33270 Floirac, France}
\email{bacmann@obs.u-bordeaux1.fr}

\and

\author{Th. Henning}
\affil{Max-Planck-Institut f\"ur Astronomie, K\"onigstuhl 17, D-69117 Heidelberg, Germany}
\email{henning@mpia.de}



\begin{abstract}
We present a ray-tracing technique 
for radiative transfer modeling of complex three-dimensional (3D) structures which
include dense regions of high optical depth like in dense molecular clouds,
circumstellar disks, envelopes of evolved stars, and dust tori around active
galactic nuclei. 
The corresponding continuum radiative transfer problem is described
and the numerical requirements for inverse 3D density and temperature
modeling are defined.
We introduce a relative intensity and transform the radiative transfer equation
along the rays to solve
machine precision problems and to relax strong gradients
in the source term. For the optically thick regions where common
ray-tracers are forced to perform small trace steps, we give
two criteria for making use of a simple approximative solver
crossing the optically thick
region quickly.
Using an example of a density structure with optical depth changes of
6 orders of magnitude and sharp temperature variations, we demonstrate
the accuracy of the proposed scheme using a common
5th-order Runge-Kutta ray-tracer with adaptive step size control.
In our test case, the gain in computational speed is about a factor of 870.
The method is applied to calculate the temperature distribution within a massive
molecular cloud core for different boundary conditions
for the radiation field.
\end{abstract}





\section{Introduction}
Almost all our information about astrophysical objects have been obtained
by analyzing their radiation. Therefore, radiative transfer (RT) modeling of
spectra and images is a standard technique of astrophysical research.
Moreover, radiation transport is an important physical process to
transport energy and momentum, often controlling the energy balance
of the object, or altering its appearance by radiation pressure
\citep[see, e.g.,][]{2002ApJ...569..846Y, 2000ApJ...539..258R}.
Given those two major applications of radiative transfer, it could be
expected that radiative transfer modeling is a well-advanced and elaborated
astrophysical
tool being established for many years already. On the contrary,
the solution of 3D radiative transfer problems is still one of the
outstanding challenges in computational astrophysics.
This is due to the high dimensionality of the radiation field (coordinates
for the location in space, for the direction, the wavelength, and the time),
the many
orders of magnitude in the change of density and spatial scale that may have to
be covered, the integro-differential character of the transport equation,
and the nonlinear coupling of the different energies ranges through the local
energy balance \citep[see, e.g.][]{2003A&A...401..405S}.
Together with the many free parameters entering complex structures, this also
has prohibited modeling of images with inverse radiative transfer, aside from
very simple cases where semi-analytical solutions can be found 
\citep{2002JQSRT..74..183S}.

The currently proposed methods to solve the continuum RT problem
are based on the Monte-Carlo approach,  finite differencing schemes on
(adaptive) grids, moment approaches, and solving the radiative transfer
equation along rays (ray-tracing) , often with a mixture
of techniques implemented into one program \citep[for a review of
radiative transfer techniques, see e.g.][]{2004A&A...417..793P, HenningAIU200,
1990ARA&A..28..303K, Hubenybuch, 1978stat.book.....M}.

A common problem of most RT codes is the treatment of the densest regions
of astrophysical objects like molecular cloud cores, accretion disks, or
dust tori around active galactic nuclei,
where the optical depths can reach high values like $10^6$ depending on the
wavelengths of the considered radiation (see Tab.~1 for typical peak values).
Applied to these objects,
Monte Carlo simulations have to calculate many interactions of the photons
\citep[for an improvement see][]{2003CoPhC.150...99W},
adaptive grid techniques refine the regions into many cells 
\citep{2002JQSRT..75..765S}, and
ray-tracer are forced to do small tracing steps. Moment methods are
well-posed to treat the optically thick regime. However, in parts of the computational
domain where the optical depth is at the order of or smaller than unity, the radiation 
field can become strongly varying with direction, and many orders of momenta 
are required to describe the radiation field correctly.
Moreover, errors in calculating the optically thick regions correctly have
little influence on the energy balance within the computational domain,
but will alter the appearance of the object strongly as the outgoing
radiation of the optically thick regions defines the inner boundary for the
layer of the object where the optical depth is of the order of unity.

Ray-tracing solvers have the advantage that the numerical error can be
controlled precisely. General purpose schemes to solve ordinary differential
equations along rays are available
in high-order accuracy and with adaptive step size control, like the advanced 5th-order
Runge-Kutta solver proposed in \citet{NR}.
The drawback of such solvers, however, is that the known functions of
the transport equations have to be evaluated very often in order to
cancel all low-order error terms. Aside from this, they are not making explicit use
of the fact that the radiation field becomes very simple when the optically
thick case is reached.

The ray-tracing follows the transport of the radiation often given on an 
underlying grid. In the short characteristics approach, the change of the
intensity is interpolated onto the local grid 
\citep[see, e.g.,][]{2006NewA...11..374M}, 
while in long characteristics 
approaches, radiation is transported between two points.

\begin{table}
\begin{center}
\caption{Typical peak optical depths ($\lambda=550$ nm) for dense dusty environments\label{tbl1}}
\begin{tabular}{lcc}
\tableline\tableline
environment & $\tau$\\
\tableline
low-mass molecular cloud core\tablenotemark{a} & 100 \\
massive molecular cloud core\tablenotemark{b} & $10^5$ \\
circumstellar disk around a young star\tablenotemark{c}& $10^4$ \\
dust torus around an active galactic nucleus\tablenotemark{d} & 100\\
\tableline
\end{tabular}
\tablenotetext{a}{derived from the optical depth at 7 $\mu$m found in \citet{2000A&A...361..555B}}
\tablenotetext{b}{\citet{2005ApJ...628L..57W}}
\tablenotetext{c}{not measurable yet, standard accretion disk model and common 
interstellar dust grains are assumed}
\tablenotetext{d}{see, e.g., \citet{2005ApJ...618..649W}}
\end{center}
\end{table}

For the case of the spectral energy distribution of a circumstellar disk around a low-mass star, 
different numerical RT techniques have been compared in \citet{2004A&A...417..793P}
to establish a benchmark for 2D continuum RT. 
A somewhat unrealistically flat radial profile with a power law index of
-1 for the density distribution and a large inner radius was assumed to reduce the 
optical depth in the disk and therefore the numerical effort.
Steepening the radial profile towards more realistic values around -2 and assuming
inner disk radii around a few stellar radii up to 10 stellar radii will increase the optical
depth in the midplane to values around $10^4$. 
The next step, therefore, is to develop numerical schemes that are able to perform
accurate calculation even in region where the optical depth is larger
than 100 \citep[maximal value in the 2D RT benchmark][]{2004A&A...417..793P}.
In this paper, we suggest a scheme for ray-tracing to overcome this problem.

We propose to transform the ray equation into a form that
is more appropriate for high-order ray-tracers applied to high-opacity
structures.
In Sect.~2, we define the radiation transport equation
and the intensity transformation as well as the modified transport
equation. We discuss the advantages of the transformation
in terms of limited machine precision and applicability of approximative
tracing in the optically thick regions.
We apply the solver to the test case 
of radiation crossing a medium with large density and
temperature gradients in Sect.~3 and compare the results
to the solution obtained with common ray-tracing.
In Sect.~4, we use the density structure from a recently performed
multi-wavelength modeling of a
complex dense molecular cloud core, rescaled to higher masses,
to investigate the thermal influence of a nearby illuminating star.
The findings are summarized in Sect.~5.

\section{Transformation of the continuum radiative transfer equation}
While the scheme proposed in this paper is applicable to both line and
continuum RT, we will introduce and demonstrate its capabilities 
by concentrating on continuum radiation.
We consider a configuration of gas and dust being exposed to radiation
from internal or external sources (for astrophysical applications,
these might be stars, nearby gas and dust etc.). The radiation may
interact with the dust particles in form of absorption or scattering,
and the dust particles are treated like black body emitters with the
temperature $T$. The stationary specific intensity $I_\lambda$ of the
radiation field is a function of the location $\vec x$, the direction
$\vec n$, and the
wavelength $\lambda$. The corresponding RT equation
reads
\begin{eqnarray} \label{rte}
\vec n \nabla_{\vec x}
I_\lambda(\lambda,\vec x,\vec n)
&=&
-\left[
  \kappa^{abs}(\lambda,\vec x)+
  \kappa^{sca}(\lambda,\vec x)
 \right]
I_\lambda(\lambda,\vec x,\vec n)
+ \kappa^{abs}(\lambda,\vec x)
B_\lambda[\lambda,T(\vec x)]\cr
&&
\frac{\kappa^{sca}(\lambda,\vec x)}{4\pi}
\int\limits_\Omega
 d\Omega'\ p(\lambda,\vec n,\vec n')
I_\lambda(\lambda,\vec x,\vec n')
\end{eqnarray}
with the phase function $p$ describing the
probability to scatter radiation from direction $\vec n'$ into $\vec n$
and the specific Planck function $B_\lambda$ at temperature $T$.
The absorption and scattering coefficients $\kappa^{abs}$ and
$\kappa^{sca}$ are defined by the product of the number density at
the location $\vec x$ and the cross section for absorption and scattering,
respectively.
Scattering of radiation at dust particles is important for short
wavelengths and depends on the size, shape and chemical composition
of the dust particles. The scattering integral in (\ref{rte}) makes
the application of an
explicit ray-tracing scheme impossible, and iterative schemes are
commonly used by fixing the intensity to a known start value
$I_0$, calculating the scattering integral, and solving
(\ref{rte}) for an updated $I_1$ until convergence is reached.
Another way is to split the
intensity into an unprocessed, a scattered, and re-emitted component
and to treat these components with different solvers \citep{2003A&A...401..405S}.
We will assume in this paper that the scattered part of the radiation
is treated separately and will concentrate on the
unprocessed and re-emitted radiation causing difficulties when
regions of high optical depth are encountered.

We trace the radiation along a ray with a given initial value
$I_\lambda(\lambda,x_0)$ at the point $x_0$
where $x$ is the coordinate along
the ray.
The solution of (\ref{rte}) in the direction $\vec n$ of the ray
for vanishing scattering is
\begin{equation} \label{solution}
I_\lambda(\lambda,x)=I_\lambda(\lambda,0)\exp\left[-\tau(\lambda,0,x)\right]
+\int\limits_{0}^{x}dx'\ \kappa^{abs}(\lambda,x')
B_\lambda[\lambda,T(x')]
\exp\left[-\tau(\lambda,x',x)\right]
\end{equation}
with the optical depth
\begin{equation}
\tau(\lambda,x_i,x_j)=\int\limits_{x_i}^{x_j}dx'\
\kappa^{abs}(\lambda,x').
\end{equation}

Introducing a typical length scale $L$ for changes in the intensity or
density structure to be considered, we define a region within the computational
domain to have a high optical depth if $\kappa^{abs}(\lambda,x)L\gg1$.
For the case of local thermal equilibrium, Eq.
(\ref{solution}) approaches the limiting case
$I_\lambda(\lambda,x)=B_\lambda[\lambda,T(x)]$
as the local thermal emission dominates the intensity due to the
strong damping of emission at preceding locations along the ray.

Although the differential equation is of simple structure and
the solution incorporates only an integration over a known
function, the numerical evaluation of (\ref{solution}) is not straightforward.
The reasons are the two exponential functions entering the radiation equation.
First, the exponential containing the optical depth has to be
calculated precisely along the ray.
Given a limited computational range of the computer, an optical depth of
1000 usually exceeds these limits and makes it necessary to renormalize the intensity.
Second, the Planck function
\begin{equation}
B_\lambda(\lambda,x)=
\frac{b(\lambda)}{e^{t(\lambda)/T(x)}-1}
\end{equation}
rises sharply with $\lambda$ for temperatures in the Wien part $T(x)<t(\lambda)$, with
$t(\lambda)=hc/\lambda k$
and $b(\lambda)=2hc^2/\lambda^5$.

A substitution of the intensity transforming the quantities
into a range where numerical limits of the computer are
not encountered and the exponential terms of the equation
can be controlled is therefore desired.

Another requirement for the ray-tracing is speed.
To model complex density structures, a parameter number of several hundreds is
usually required \citep[see, e.g.,][]{2005A&A...434..167S}. Given a certain parameter
set describing the density structure, 
the temperature within the object is determined and used
to calculate theoretical images at different wavelengths. These images are
compared with the observed images and a minimization tool like simulated
annealing is used to optimize the model parameters. Assuming several thousand
pixels to be fitted simultaneously, and a minimum of $10^4$ steps of the
optimization tool, the total number of rays often exceeds $10^8$. To perform
the calculation within a reasonable time, the ray-tracer should be optimized
with respect to speed.

An advanced ray-tracer with 5th-order accuracy has been proposed by
\citet{NR}, based on the Runge-Kutta scheme. It is coupled with
an adaptive step size control using an embedded form
of the 4th-order Runge-Kutta formula. As parameters for the error
truncation, values determined by \citet{1990Cash} are used.

The disadvantage of this tracer is that the density and the temperature
function are called six times each in every tracer step. Each of the called
functions contains several hundred parameters for a complex 3D structure
making an evaluation expensive.
On the other side, the tracer steps are chosen adaptively and with full error
control. The tracer is therefore the first choice for astrophysical problems
with moderate optical depths variations.

A major numerical difficulty arises for regions with high optical depth
$\kappa^{abs}(\lambda,x)L>100$.
Neglecting scattering, the change in the intensity expressed in the right-hand side
of (\ref{rte}) is proportional to $\kappa^{abs}(\lambda,x)$.
Having calculated the intensity at a point $\vec x$, the difference
$|B_\lambda(\lambda,x)-I_\lambda(\lambda,x)|$ is small as the
local thermal emission dominates the radiation field in the
optically thick regime. A small change in the intensity suggests that the
solver is able to perform large spatial steps. However, the difference
$|B_\lambda(\lambda,x)-I_\lambda(\lambda,x)|$ is multiplied with the large absorption 
term $\kappa^{abs}(\lambda,x)$ amplifying any
change in the source term that is arising from the spatial variation in the temperature.
To meet the accuracy requirements, the tracer therefore performs small steps
although the change in the intensity is known in the optically thick
limit and the local source term has no strong gradients.
A modification of the tracer, therefore,
would be desirable in order to both crossing the regions of high optical depths
quickly without
introducing large errors and to keep the solution away from the floating point limits.

In order to develop a ray-tracer meeting those requirements and being acceptable
in terms of computational effort,
we introduce the relative intensity
\begin{equation}\label{relint}
D_\lambda(\lambda,x)\equiv
\frac{I_\lambda(\lambda,x)}
{I_\lambda(\lambda,x)+B_\lambda(\lambda,x)}.
\end{equation}
For a vanishing source function, it approaches unity, and for the
optically thick part with $I_\lambda(\lambda,x)\approx B_\lambda(\lambda,x)$,
$D_\lambda(\lambda,x)$ has a value close to $1/2$.
Inserted into (\ref{rte}) and neglecting scattering, we obtain the differential equation
\begin{equation}
\frac{dD}{dx}=
\left[1-D_\lambda(\lambda,x)\right]
\left(
\kappa^{abs}(\lambda,x)\left[1-2D_\lambda(\lambda,x)\right]-
\frac{D_\lambda(\lambda,x)}{B_\lambda(\lambda,x)}
\frac{dB_\lambda(\lambda,x)}{dx}
\right).
\end{equation}
The relative intensity is chosen in a way that the source function in the
differential equation with
the exponential dominating the Wien approximation regime
is transformed to the function
\begin{equation}
\frac{1}{B_\lambda(\lambda,x)}
\frac{dB_\lambda(\lambda,x)}{dx}
=\frac{t(\lambda)}{T(x)^2}
\frac{dT}{dx}
\frac{1}{1-e^{-t(\lambda)/T}}.
\end{equation}
The denominator $1-\exp[-t(\lambda)/T]$ has a value close to 1 for $t(\lambda)\gg T$
so that the exponential terms no longer cause numerical problems.

Another question is if the 
used rational function transformation amplifies numerical errors.
However, no strong solution error amplification is introduced by transforming to the
relative intensity, as the relative error in $D_\lambda(\lambda,x)$
\begin{equation}
\frac{\pm\Delta D_\lambda(\lambda,x)}{D_\lambda(\lambda,x)}=
\frac{\pm\Delta I_\lambda(\lambda,x)}{I_\lambda(\lambda,x)}
\frac{2I_\lambda(\lambda,x)+B_\lambda(\lambda,x)}
{I_\lambda(\lambda,x)\mp\Delta I_\lambda(\lambda,x)+B_\lambda(\lambda,x)}
\leq
\frac{\pm 2\Delta I_\lambda(\lambda,x)}{I_\lambda(\lambda,x)\mp
\Delta I_\lambda(\lambda,x)}
\end{equation}
is at most only about a factor 2 larger than the error in the intensity if
$\Delta I_\lambda(\lambda,x)\ll I_\lambda(\lambda,x)$.

To obtain a criterion for the use of an approximate solver in the optically
thick region $D_\lambda(\lambda,x)\approx 1/2$, we introduce a positive limit
$\varepsilon$ for the relative
difference of intensity and Planck function with
\begin{equation}\label{epsi}
\frac{|B_\lambda(\lambda,x)-I_\lambda(\lambda,x)|}
{B_\lambda(\lambda,x)}=\frac{|1-2D_\lambda(\lambda,x)|}{1-D_\lambda(\lambda,x)}<\varepsilon.
\end{equation}
Using 1st-order finite differencing for (\ref{rte}) for two neighboring points
$x_1$ and $x_2$ and neglecting scattering, 
the limiting case of condition (\ref{epsi}) reads
\begin{equation}
\frac{
 \frac
   {|\Delta B_\lambda(\lambda,x)|}
   {B_\lambda(\lambda,x)}
     }
     {\Delta x \kappa^{abs}(\lambda,x)}
 =
\pm\varepsilon
\end{equation}
with $\Delta x=x_2-x_1$ and
$\Delta B_\lambda(\lambda,x)=B_\lambda(\lambda,x_2)-B_\lambda(\lambda,x_1)$.
The sign in front of $\varepsilon$ refers to the sign of
$B_\lambda(\lambda,x)-I_\lambda(\lambda,x)$.
The condition is fulfilled either due to small relative changes in the
source term $|\Delta B_\lambda(\lambda,x)|/B_\lambda(\lambda,x)$ 
or a large local optical depth $\Delta x \kappa^{abs}(\lambda,x)$.
In a pre-calculation along the ray, the regions where the condition is fulfilled 
can be identified. Then, a fast solution can be obtained for these region by
performing large steps while assuming $D_\lambda(\lambda,x)=1/2$.
As the limit for validity, we have used $\varepsilon=10^{-6}$ in this paper.

\section{Application of the ray-tracing scheme to
a configuration with $1<\tau<10^6$}
To test the proposed ray-tracer, we choose a configuration with
variations in the local optical depth covering six orders of magnitude.
The temperature profile is chosen so that it includes both external
heating and a deeply embedded radiation source. Such a configuration is
often found, e.g., in star formation regions where young stars illuminate
clouds where others embedded star are currently forming.

As density profile, a sum over two Gaussian distributions is used for the
number density of the absorbing material (representing, e.g., two molecular
clouds containing gas and dust)
\begin{equation} \label{ndens}
n(z)=n_1\exp\left[-\left(\frac{z-z_1}{w_1}\right)^2\right]
+n_2\exp\left[-\left(\frac{z-z_2}{w_2}\right)^2\right]
\end{equation}
with the two optical depths ($i=1,2$)
$\tau_i(\lambda,x)=\kappa_i(\lambda,x) L=\sigma^{abs}(\lambda) n_i(z) L$
having the values $10^6$ and $10^4$, respectively, the
absorption cross section $\sigma^{abs}(\lambda)$, and introducing the
dimensionless spatial variable $z=x/L$. The widths of the Gaussians are
chosen as $w_1=1/2$ and $w_2=1$.
We will calculate the intensity along a representative ray that crosses both density maxima and
encounters strong variations in the source term. In
a real application, of course, all rays will be calculated (see Sect.~4).

Without calculating the correct temperature reached in local thermal
equilibrium for this test case, we assume a temperature distribution
close to what can be expected for the density distribution defined in
(\ref{ndens})
\begin{equation} \label{temp}
T(z)=T_0 (z+z_0)^{-1/2}+T_3\exp\left[-\left(\frac{z-z_3}{w_3}\right)^2\right]
+T_4\exp\left[-\left(\frac{z-z_4}{w_4}\right)^2\right] + T_{bg}.
\end{equation}
The first term represents a temperature profile arising from an
external heating by a distant radiation source (e.g. a star) located in an optically thin
region at large negative $z$ ($T_0=50$ K, $z_0=L/2$).
Numerically, this corresponds to
an exponential drop of the intensity that would quickly reach the
machine precision limit without a source term. Therefore, the
source term contribution, although being low, is determining
the intensity in the center of the clump and saves the solver
from failing due to precision problems.

An intermediate region is reached between the two clumps where
external radiation
can illuminate the material from other directions.
Therefore, we have chosen the temperature to have a broad Gaussian profile
with $T_3=10$ K, $z_3=3.7L$, and $w_3=L$, peaking where the density
has a minimum.

The second clump is less dense, but hosts an embedded source of heat
represented by the second compact Gaussian profile in (\ref{temp}) with
$T_4=1500$ K, $z_4=6L$, and $w_4=L/10$. In this case, the tracer
enters a region where the source distribution rises sharply and
dominates the radiation field again.
A constant background radiation field in the far-infrared wavelength
region will heat up the matter slightly, represented in our test case
by the constant background temperature $T_{bg}=5$ K.

\begin{figure}[htp]
\centering
\includegraphics[width=15cm]{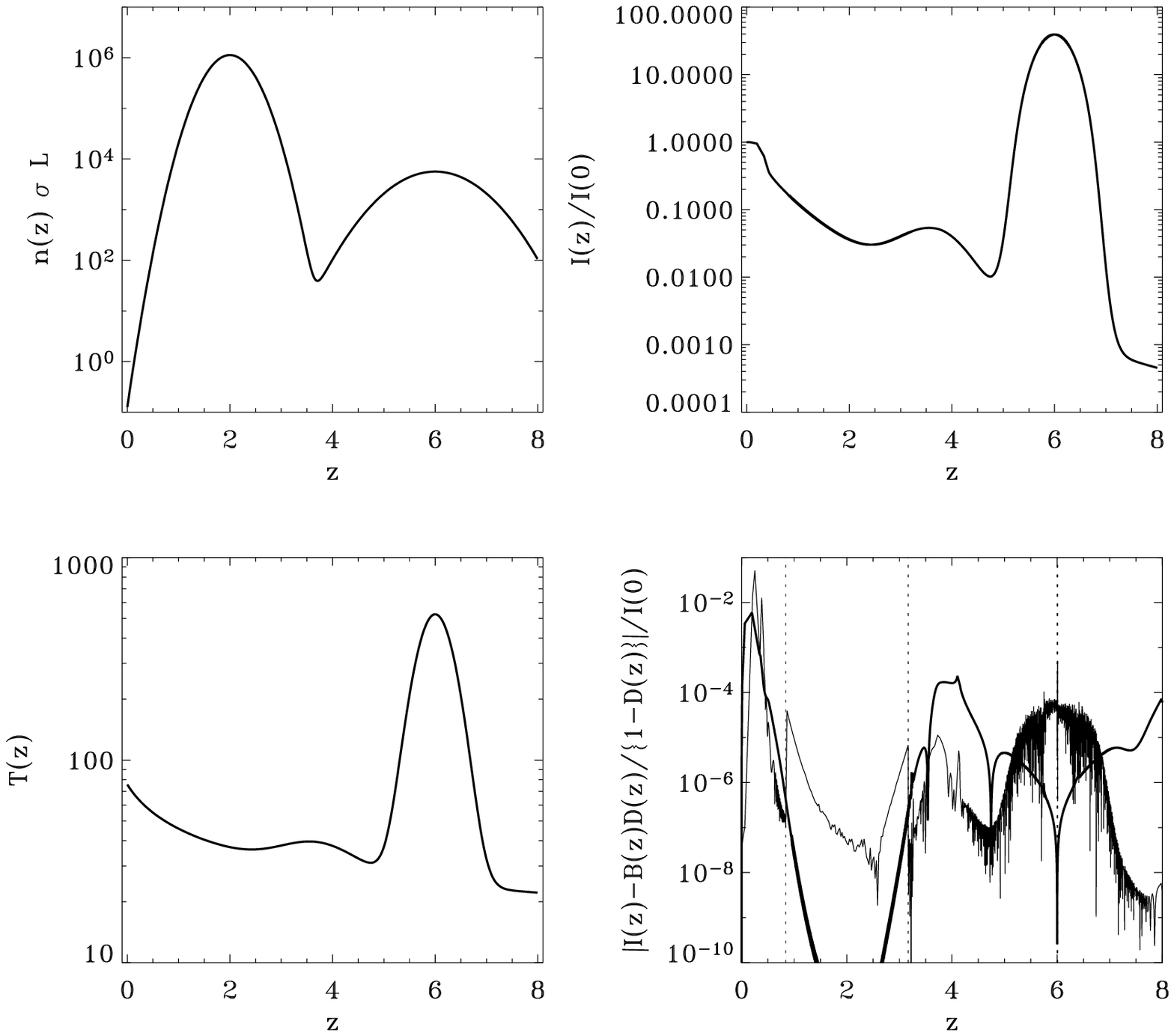}
\caption{
Overview of the physical set-up and solutions for the test case
calculation. Top left: Optical depth 
along the ray.
Bottom left: Temperature along the ray.
Top right: Intensity along the ray using a 5th-order Runge-Kutta
solver for the ray equation.
Bottom right: Normalized difference between the solutions obtained
from the accurate solver and 
the relative intensity ray-tracer (thin line) and interpolation
error (thick line).
Regions where the 
solver uses the optically thick approximation are enclosed by vertical dotted
lines (z=0.9 to z=3.2, and around z=6).
}\label{fig:1}
\end{figure}

We have chosen a wavelength of $\lambda=6\ \mu$m for this test case
because the thermal
emission defined by the temperature profile defined in (\ref{temp}) gives
a substantial contribution to the radiation field. Note that in
common star formation regions, the local optical depth will
reach $10^6$ only in the densest regions at this wavelength.
Fig.~\ref{fig:1} gives an overview over the test case and the
numerical solution of the corresponding radiative transfer equation.
The upper left panel shows the local optical depth
$L \sigma^{abs}(\lambda) n(z)$ as a function of $z$ with the two Gaussian
profiles establishing a density change of up to six orders of magnitude.
In the lower left panel, the temperature variation along the ray is given,
with a slow decrease in the outer parts of the first density clump, a
local maximum between the clumps, and a strong bump in the center
of the second density maximum.

Applying the Runge-Kutta scheme with an solution accuracy of $10^{-6}$
and $\sim$213000 main tracing steps to the entire region,
we show in the upper right panel
the numerical solution of the radiative transport equation
$I^{full}_\lambda(\lambda,z)/I_\lambda(\lambda,0)$ along
the ray. The normalized intensity decreases
exponentially from the start value 1, entering a region where the
thermal emission dominates the
solution at $z\sim 0.4$. The shape of the intensity
reproduces the temperature side maximum at $z=3.8$.
At $z\sim 4.8$, the Planck function rises quickly due to the thermal
peak centered at $z=6$. The Gaussian shape in the intensity is broader
than the corresponding temperature profile as the optical depth is $\tau<10^4$
so that a small contribution of preceding radiation is added to the local
thermal emission in the source integral of (\ref{solution}). An additional 
factor for the broadening is that
for low temperatures $T(z)<t$, the Planck function rises as
$\sim\exp[-t/T(z)]$, while around the peak at $z=6$ with $t\approx$T(z)
it is turning to follow the $B_\lambda(\lambda,z)\sim T(z)$-dependence 
for the large temperatures.

The solution of the transport equation using the
proposed tracer for the relative intensity
$D_\lambda(\lambda,z)$, transformed back to an intensity $I^{rel}_\lambda(\lambda,z)$
using
(\ref{relint}), is overlayed in the upper right panel and can not be
distinguished from $I^{full}_\lambda(\lambda,z)$.

The absolute value of the difference between the full solution
and the relative solution
\begin{equation}
\left| I_{full}-I_{rel}\right|=
\left| I_\lambda(\lambda,z)-
\frac{D_\lambda(\lambda,z)B_\lambda(\lambda,z)}{1-D_\lambda(\lambda,z)}
\right|
\end{equation}
is shown in the lower right panel, normalized to the start intensity (thin line).
The dotted vertical lines at $z\approx 0.8$ and $z\approx 3.2$
indicate where the relative tracer has switched to an accelerated crossing
of the optically region and back to full tracing, respectively.

Before discussing the solution error, it should be considered that
calculating the difference of the solutions
requires to interpolate one solution onto the other, as they have
been obtained using different
adaptive step sizes.
This introduces a new interpolation error into the difference that has
to be distinguished from the error related to the solution process.
To estimate the order of magnitude interpolation error,
we have overplotted in the lower right panel the
absolute value of the error of a
linear interpolation of the full normalized intensity onto the z-values
of the relative intensity values as a thick line.
The second derivative of $I_\lambda(\lambda,z)$ can not be used, as the
full solution is only known at discrete values, so that calculating the
second derivative would introduce new interpolation errors into the error analysis.
We therefore estimate the error in the optically thick
regions with $D_\lambda(\lambda,z)\approx 1/2$
to be
\begin{equation}
\varepsilon_{lim}=
\left.
\frac{\Delta z^2}{2I_\lambda(\lambda,0)}
\frac{\partial^2 B_\lambda(\lambda,z)}{\partial z^2}\right|_{z_0}
\end{equation}
with the second derivative of the Planck function being
\begin{equation}
\frac{\partial^2 B_\lambda(\lambda,z)}{\partial z^2}=
\frac{\partial B_\lambda(\lambda,z)}{\partial z}
\left[
\frac{1+e^{-t(\lambda)/T(z)}}{B_\lambda(\lambda,z)}
\frac{\partial B_\lambda(\lambda,z)}{\partial z}
-\frac{2}{T(z)}
\frac{dT(z)}{dz}
+\frac{d^2T(z)}{dz^2}
\left(
\frac{dT(z)}{dz}
\right)^{-1}
\right].
\end{equation}
The interpolation error shown as a thick line in the lower right panel starts
at a high value of $10^{-2}$ as the step size of the full tracer is large
in the region where the density is low and the thermal contributions
are small.
The error becomes very small in the region of high optical
depth as the full tracer is enforced to use small steps entering
the interpolation error quadratically. The interpolation error has
roots wherever the temperature has extrema as
$\partial^2 B_\lambda(\lambda,z)/\partial z^2\sim dT(z)/dz$.

For small $z$, the plotted solution difference is of the order of the
interpolation error, and the real difference is likely
to be much smaller as both solutions are obtained by the 5th-order
Runge-Kutta scheme in this region.
In the optically thick region between the first two dotted lines,
the interpolation error is below $10^{-7}$, and the thin curve shows the difference
between the full scheme and the rapid cross stepping used by the
relative solver. Still, this error is small and reveals the real
benefit of the accelerated tracer.
While the full solver spends 209000 grid points to trace the thick
region, the relative solver uses 240 steps only with a difference
in the solution that is of the order of $10^{-4}$ to $10^{-6}$.

A second pair of dotted lines appears at $z\approx 6$, but the interval
is so small that it appears as one line.
The reason that the relative solver is not switching to larger steps
although the local optical depth is $10^4$ for the second
density maximum is related to the large gradient in the source function.
Only in the maximum of the source term, the change in the source
function becomes small enough so that the change in the relative
intensity allows large stepping for a small range.
The difference in the two full solutions is up to
an order of magnitude larger than the
$10^{-6}$-accuracy limit of the 5th-order Runge-Kutta scheme, as
the error in D is twice the error in I, and interpolation errors
and tracer errors are added up in the thin line.

To summarize the findings for this particular test case,
the relative tracer can reproduce the solution achieved
by the full tracer with an accuracy of about $10^{-5}$ while it
crosses the optically very thick regions 870 times faster than the full tracer.
The relative tracer is sensitive to changes in the source function
to achieve this accuracy even in cases where strong intensity
changes are caused by a varying source function within optically
thick regions.
The concrete gain in performance when using the relative tracer will depend
on the extent and optical depth value of the dense regions to be
crossed as well as on the occurrence of complications like embedded
sources.

\begin{figure}[htp]
\centering
\includegraphics[width=15cm]{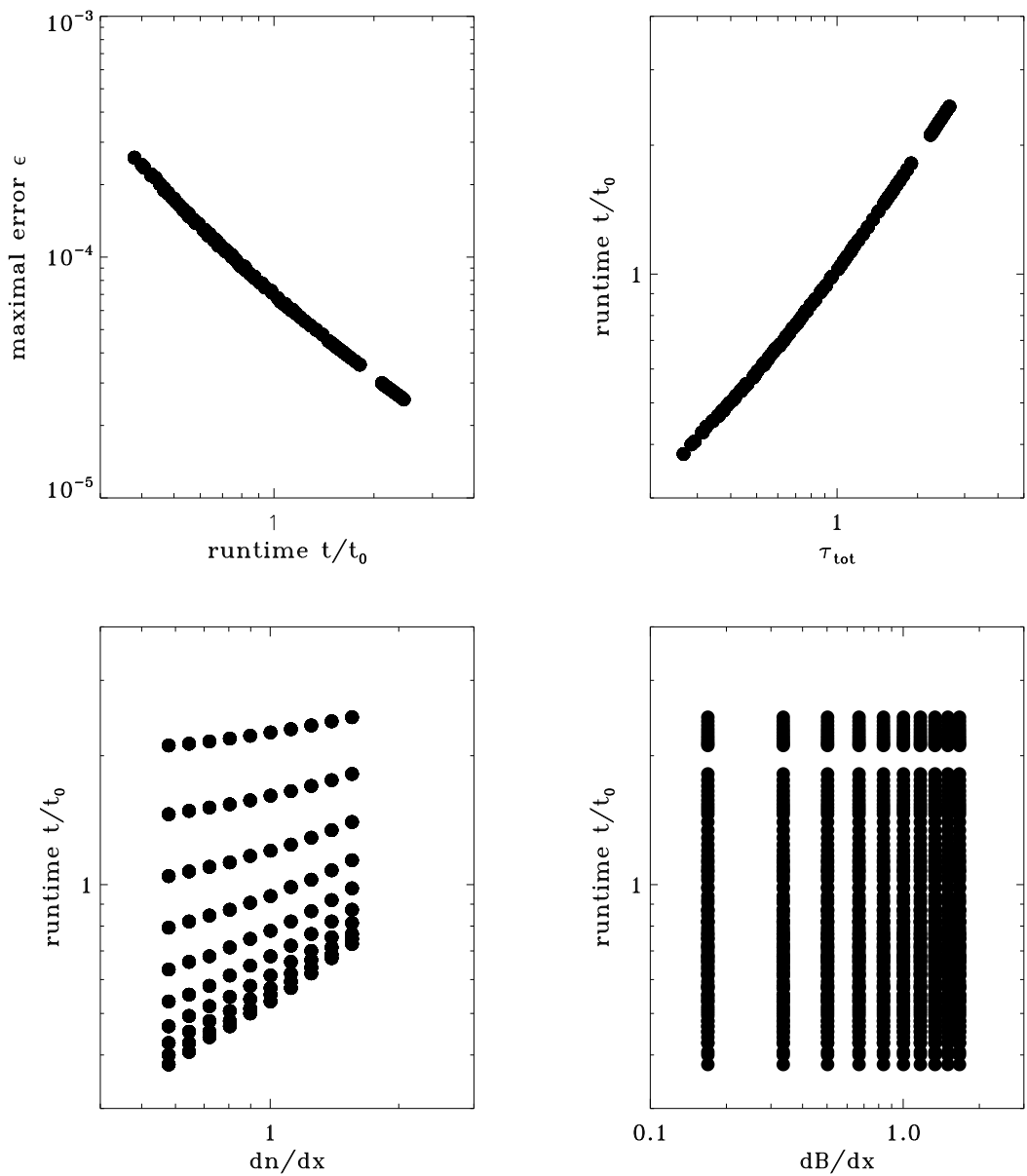}
\caption{
Top left: Relation between the run time and the maximal error between
relative solution and accurate solution.
Top right: Run time as a function of the total optical depth for a ray 
crossing the region.
Bottom left: Run time as a function of the density gradient maximum.
Bottom right: Run time as a function of the maximal gradient in the
source function.
All quantities aside from the maximal error are normalized to the test
case shown in Fig.~\ref{fig:1}.
}\label{fig:2}
\end{figure}

We have performed 1000 test runs with varying total optical depth,
density gradient, and source function gradient to explore the overall
influence of the dominant ingredients on the run time and the error
maximum.
The resulting dependencies are shown in Fig.~\ref{fig:2}. 
The top right
panel shows the relation between the total optical depth for a ray
crossing the considered region and the run time normalized to the test
case discussed in this section. They are directly proportional, as the
adaptive steps of the Runge-Kutta solver scale inversely
with the optical depth. For higher optical depth, the approximation
of the relative tracer is better fulfilled, decreasing the maximal error
for higher total optical depth. Therefore, the maximal error decreases
inversely with the run time, as shown in the top left panel.
The dependence on the density gradient is depicted in the lower left panel.
For small run times, the run time rises linearly with maximal density gradient.
This rise gets weaker for larger run times, as most of the run time is
taken by the overall high optical depth. The lower right panel shows
that there is no dependency of the run time on the gradient in the
source function as the large optical depth dominates the run time
in the region where the relative intensity solver is used.

\section{Application of the ray-tracing scheme to
3D illumination effects of a dense molecular
cloud core}
To demonstrate the capabilities of the new
ray-tracing scheme, we apply the scheme to a more realistic astrophysical
problem, incorporating a large number of ray-tracing calculations
through a high-opacity
massive molecular cloud core.
These
cores are of particular interest as they
are thought to be the progenitors of young stars.
Characterizing their physical conditions will lead to a better
understanding of how stars are forming in molecular clouds.
The detailed understanding of images of cores crucially relies
on the knowledge of the 3D structure, as projection effects
limit the reliability of simple 1D modeling \citep[e.g.,][]{2004ApJ...615L.157S}.
In \citet{2005A&A...434..167S}, we have fitted images of the
cloud core $\rho$ Oph D
in the star formation region Rho Ophiuchi that have been obtained
at mid-infrared (MIR) and mm-wavelengths. The complex elongated
structure was modeled
using a series of 30 3D Gaussian distributions, determining both
the density and temperature structure by inverse multi-wavelength
3D continuum radiative transfer.

The density structure deviates significantly from spherical
symmetry and has several local maxima. Hence, it is well-suited
to serve both as a realistic test application for the
ray-tracing scheme and as a representative shape of a dense core.
The total core mass of 2.3 solar masses, however,
is not sufficient to reach substantial optical depth ($>$10) at MIR
wavelengths. We therefore increase the core density uniformly
by a factor of 100
until the optical depth ranges from 0.1 to 1000, placing the core
among the most massive dense cores.
The temperature structure for the high mass core is
very different from the distribution determined for moderate
optical depths and an outer radiation field of a low-mass
star formation site like Rho Ophiuchi.
It depends on the radiation field
produced by nearby stars, which is most likely dominated by
the massive stars in the considered star-forming region.
The strong pressure of their emitted radiation
and the stellar wind alter the structure of the parental
molecular cloud substantially by opening up gaps and compressing
the cloud material.

In this section, we assume that the cloud core is still
surrounded by a diffuse molecular cloud
with an optical depth of 1 at $\lambda=550$ nm, and that a nearby
B6 star at a distance of 1 pc is illuminating the core.
The strong UV radiation of the star can not penetrate the
parent cloud and is absorbed
by gas, dust particles, and polyaromatic
hydrocarbons (PAHs) in a layer called a photon dominated region (PDR)
\citep[see, e.g.,][and references therein]{1996A&A...313..633V}.
The chemical and radiative processes along
with the gas dynamics are complex and definitely beyond the scope
of this paper. Here, we want to adopt a very simple model for such
a PDR region and concentrate on the treatment of the continuum radiation.
Assuming that all kinematic effects of the radiation and of a possible
wind are influencing only the outer layers of the parent cloud, 
still the long-wavelength
part of the radiation is able to reach the dense core deep in the cloud and heat it.
The gas atoms, molecules, and dust particles convert the UV radiation to
radiation at longer wavelength which propagates deeper into the cloud and the dense core
before being absorbed and re-emitted at far-infrared wavelengths
\citep{1999A&A...344..342L}.
For this test application, we will neglect the possible influence
of a shock front that is formed by the radiation ram pressure
and serves as an additional source of energy. We also neglect
heating from molecular lines other than PAH emission.
The density and temperature of the dense core is discretized
on a 3D grid with $N=64 000$ grid cells.
 To calculate the incoming
energy density of the stellar radiation, $N$ traces have to be
performed to derive the column density from the star to the
cell. Then, in each cell, the stellar intensity
of the cells can be determined
for all wavelengths.

The PDR is discretized by a layer with
$N_{PDR}\times N_{PDR}\times 1$ cells at a distance of 0.5 pc covering
a cloud surface of 0.25 pc$^2$.
The emission of dust particles can be treated as black
body radiation for particles with a size of $a>0.1\ \mu$m, making
it possible to calculate the energy balance for the dust particles
in each cell quickly. Smaller dust particles and PAHs absorb
and re-emit the radiation time-dependently, substantially increasing
a numerical treatment. For the purpose of this test calculation,
we will therefore assume that the conversion factor of stellar energy from
the UV to optical wavelengths
by the PDR is a known value $\eta$ and parameterize the PDR spectrum
by a power-law in wavelength of the form
$\lambda F_\lambda\sim \lambda^{-0.4}$ between 5 $\mu$m and
18 $\mu$m \citep{2001A&A...376..650Z}. To calculate the density of the radiative
energy absorbed in the cells, $N_{PDR}^2\times N$ traces have to be
performed (we have chosen $N_{PDR}=40$).
The interstellar radiation field will mainly contribute to the
heating in the far-infrared (FIR). Assuming it to illuminate
the configuration isotropically, we have to perform traces
covering all directions. As direction
grid on the unit sphere, we use an equally-spaced
distribution of $N_d=200$ points optimized for integration purposes
using simulated annealing \citep{1996JQSRT..56..97S}.
The spectrum proposed by \citet{2001A&A...376..650Z} was used for the interstellar
radiation field. In total, $N_d\times N$ traces have to be performed
to determine the heating by the interstellar radiation field.
The density at any location was calculated from the grid using 1st-order interpolation.

Finally, to obtain a self-consistent temperature distribution,
the cell-cell illumination has to be considered incorporating
$N^2/2-N\times N_i $ traces for the column density between the cells,
where $N_i$ is the number of iteration to reach convergence of the
cell temperatures. For cores, the self-heating is restricted to the
FIR- and mm-wavelengths range as the dust particles have low temperatures
around 20 K. This contribution to the cell temperature is low, though, since
the absorption and re-emission coefficient $\kappa^{abs}(\lambda,\vec x)$
decreases as $~\lambda^{-2}$ in this region. Thus, the temperature converges to a solution after only
a few iterations.
The total amount of traces for this example adds up to
$2.2\times 10^9$ encountering integrated optical depth up to
$10^4$. We have used the optical dust properties of Silicate particles of
size 0.1 $\mu$m proposed in \citet{1984ApJ...285...89D}
in this calculation.

\begin{figure}[htp]
\centering
\includegraphics[width=15cm]{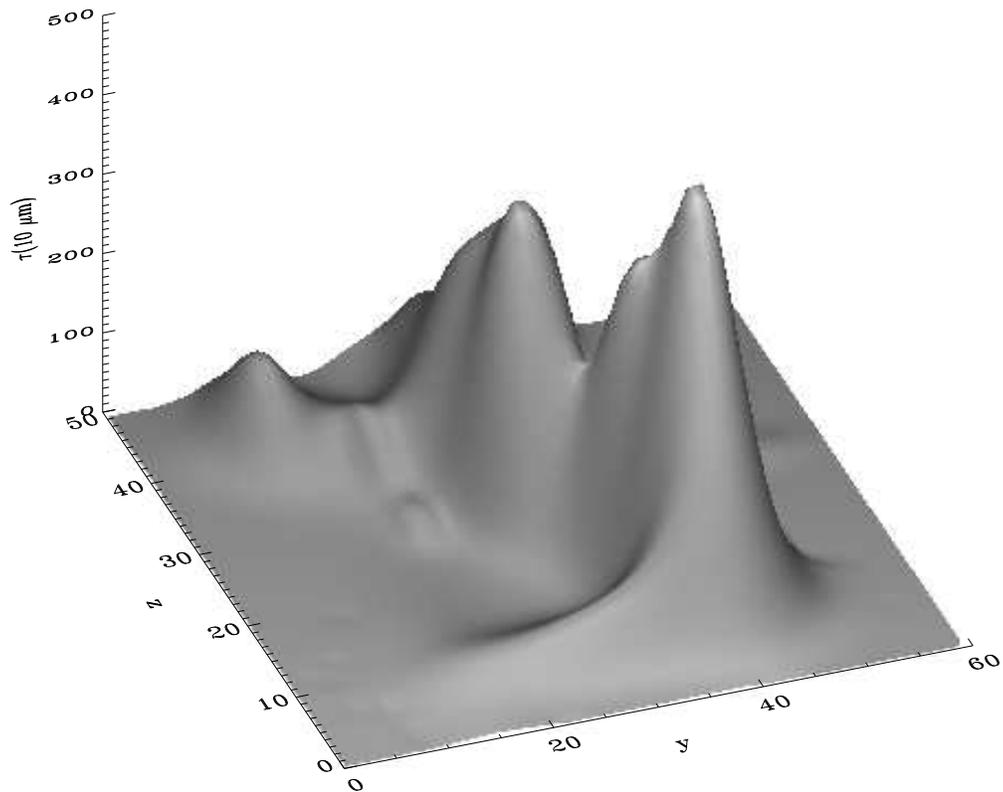}
\caption{Optical depth of a massive molecular cloud core with a density
structure adopted from a multi-wavelength 3D radiative transfer
modeling of the low-mass core $\rho$ Oph D at $\lambda=10\ \mu$m along the line of sight
(x-axis). The local plane-of-sky-coordinates are arbitrary. The
optical depth reaches values around 500.
}\label{fig:3}
\end{figure}

\begin{figure}[htp]
\centering
\includegraphics[width=15cm]{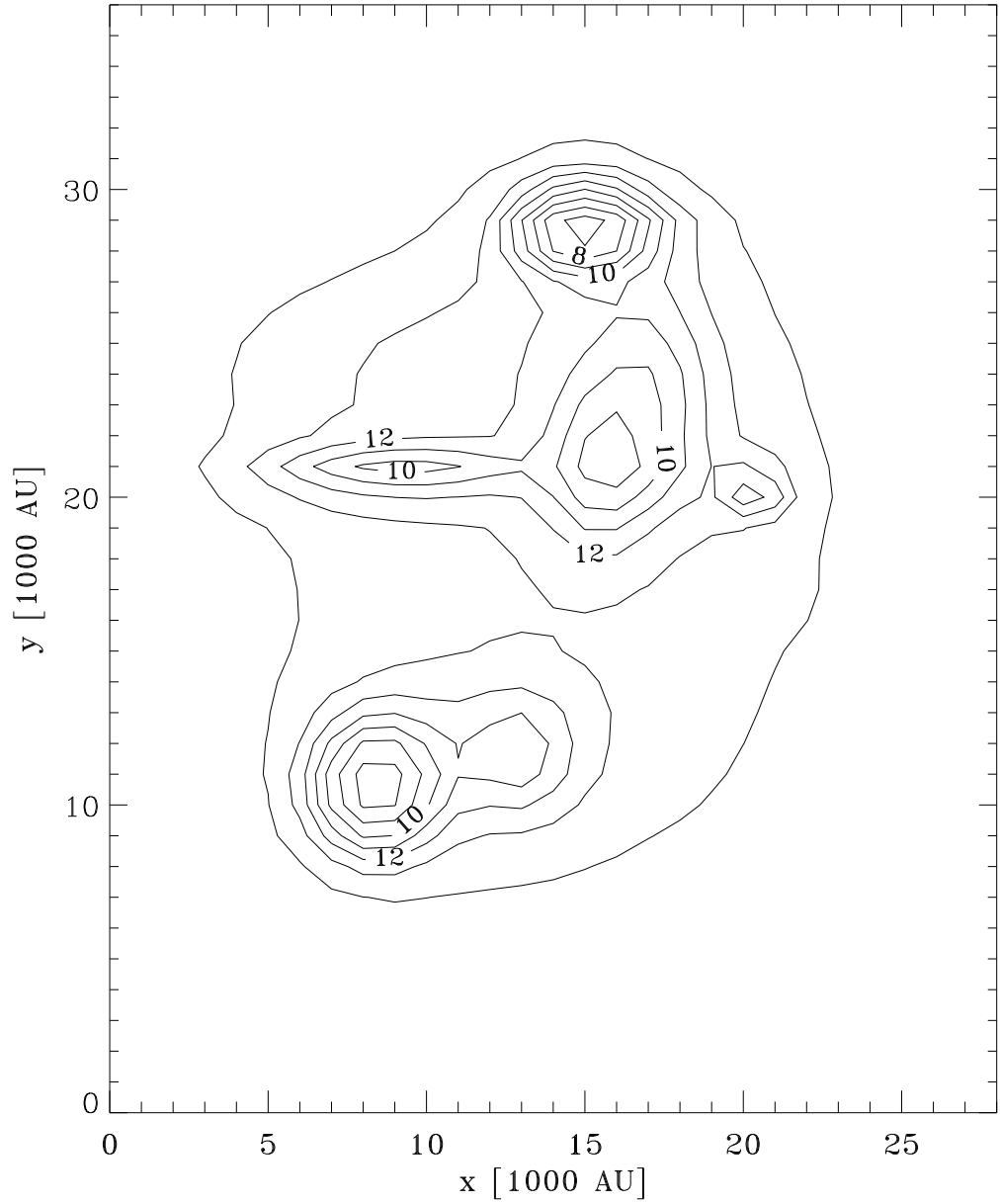}
\caption{Iso-temperature distribution within the cloud core in a plane
parallel to the plane of sky that includes the temperature minimum.
The core is heated by the interstellar radiation field only. The numbers
give the temperature in K with a 1 K stepping in the contours.
The local plane-of-sky-coordinates are arbitrary.
}\label{fig:4}
\end{figure}

\begin{figure}[htp]
\centering
\includegraphics[width=15cm]{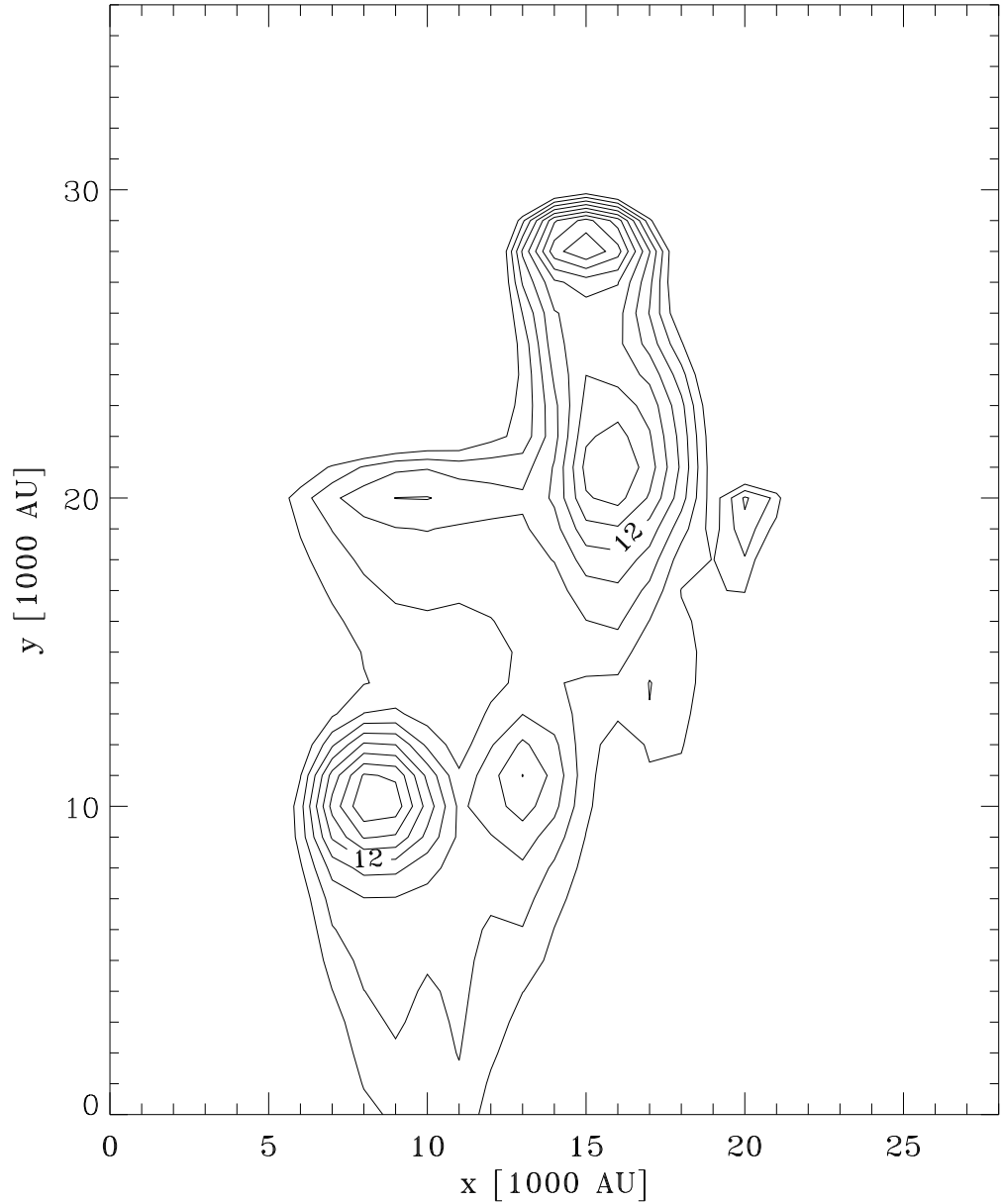}
\caption{Iso-temperature distribution within the cloud core in a plane
defined in Fig.~\ref{fig:4}.
The core is heated by a nearby B6 star, a corresponding PDR,
and the interstellar radiation field. The numbers
give the temperature in K with a 1 K stepping in the contours.
The local plane-of-sky-coordinates are arbitrary.
In this orientation, the star and PDR are located above the picture.
}\label{fig:5}
\end{figure}

Fig.~\ref{fig:3} shows the optical depth $\tau(10\ \mu$m$)$
for rays crossing the core parallel to the x-axis (y and z are given
in arbitrary coordinates in the plane of sky). The optical depth
reaches values around 500 in the densest parts.
The resulting temperature structure is visualized in Fig.~\ref{fig:4} for the
case of external heating by the interstellar radiation field. The contour
plot shows a cut through the temperature cube in a plane parallel to the plane of sky
including the point with the temperature minimum.
In massive star formation regions, this is the
case only when no massive star has formed in the vicinity yet.
Fig.~\ref{fig:5} gives the temperature in the same plane when a nearby B6 star
and the corresponding PDR is present. While the isotropic interstellar illumination
yields a temperature structure that is similar to the density structure,
the beamed illumination by the star and the PDR causes shadowing effects
with low temperature behind density maxima.
This effect will be important for the collapse of the core due to self-gravity
as the local instability criterion depends on the temperature. The influence of
external heating by nearby stars will
be investigated further in a forthcoming paper.

\section{Summary}
There is a variety of astrophysical applications requiring the calculation
of the radiation field in complex structures with regions of high
optical depth. Current radiative transfer codes suffer from the high
computational cost to determine the radiation field for such cases
as the existing solution algorithms are often not suited for
high optical depths.
In this publication, we have suggested a ray-tracing scheme
designed for the analysis of complex and dense structures based on 3D
radiative transfer. Analyzing accuracy, numerical effort,
and stepsize for the example of continuum radiation, we have found it to be well-suited for applications
including regions of arbitrary high optical depth and a vast number
of required ray-traces.
For the test case being investigated, the tracer was about 870
times faster than the full 5th-order Runge-Kutta tracer with a loss
of only one order of magnitude in accuracy.
We have applied the scheme to a
massive molecular cloud core illuminated
by a nearby massive star. Using a simple model for a photon-dominated region
at the surface of the parent molecular cloud, we have compared the 
3D temperature structure within the core for two cases: (i) an
illumination by the interstellar
radiation field and (ii) an illumination by a nearby massive star and a
corresponding photon-dominated region. Stellar illumination produces shadowed regions
of colder gas and dust which will influence the evolution of the core
possibly forming stars.

\acknowledgments
J.S. thanks the Institute for Pure and Applied Mathematics, UCLA, and
the Observatoire de Bordeaux, L3AB, for
support and beneficial discussions during visits in 2005.

\clearpage


\clearpage
\clearpage

\clearpage

\clearpage

\clearpage

\end{document}